\documentclass[aps,pre,twocolumn,superscriptaddress]{revtex4-2}
\usepackage{amssymb}
\usepackage{amsmath,bm}
\usepackage{graphicx,color}
\usepackage{bbm}
\usepackage[bottom]{footmisc}
\usepackage{multirow}
\usepackage{xcolor}
\usepackage{xpatch} 
\makeatletter   
\usepackage{xpatch} 

\newcommand{\B}[1]{{\bm{#1}}}

\newcommand{\rin}{r_\text{in}}
\newcommand{\rout}{r_\text{out}}

\newcommand{\Pvec}{\mathbf{P}}

\begin{document}
\title{Direct Measurement of Dipoles in Anomalous Elasticity of Amorphous Solids}
\author{Bhanu Prasad Bhowmik}
\affiliation{Dept. of Chemical Physics, The Weizmann Institute of Science, Rehovot 76100, Israel}
\author{Michael Moshe}
\affiliation{Racah Institute of Physics, The Hebrew University of Jerusalem, Jerusalem, Israel 9190}
\author{Itamar Procaccia$^*$} 
\affiliation{Dept. of Chemical Physics, The Weizmann Institute of Science, Rehovot 76100, Israel, $^*$Center for OPTical IMagery Analysis and Learning, Northwestern Polytechnical University, Xi'an, 710072 China}

\begin{abstract}
Recent progress in studying the physics of amorphous solids has revealed that mechanical strains can be strongly screened by the formation of plastic events that are typically quadrupolar in nature. The theory stipulated that gradients in the density of the quadrupoles act as emergent dipole sources, leading to strong screening and to qualitative changes in the mechanical response, as seen for example in the displacement field. In this Letter we firstly offer direct measurements of the dipole field, independently of any theoretical assumptions, and secondly we demonstrate detailed agreement with the recently proposed theory. These two goals are achieved using data from both simulations and experiments. Finally we show how measurements of the dipole fields pinpoint the theory parameters that determine
the profile of the displacement field.  
\end{abstract}
\maketitle

{\bf Introduction:} Classical elasticity theory is one of the basic building blocks of Condensed Matter Physics. Once the elastic moduli of a given material are known \cite{89Lutsko}, one can predict the displacement field that is associated with any mechanical strain \cite{Landau}, using linear elasticity when the latter is small, and applying nonlinear extensions of the theory when the strains are larger \cite{01CH}.
This happy state of affairs is however questioned when one studies the mechanical properties of amorphous solids. One reason is that in amorphous solids {\em plastic responses} appear instantaneously for any amount of strain \cite{10KLP}. Another reason is that nonlinear elastic constants were shown to have unbounded sample to sample fluctuations in the thermodynamic limit \cite{11HKLP}.
Finally, at least in frictional amorphous solids, one observes stress correlations
that are not consistent with classical elasticity theory \cite{21LMPR,21LMPRWZ}.
It is thus necessary to revisit elasticity theory in the context of amorphous solids with the aim of seeking an applicable theory that can provide predictive power. 

Such a theory  was presented in a recent paper \cite{21LMMPRS}. Attention was given to the plastic responses, which typically appear as quadrupolar (Eshelby-like) irreversible responses \cite{54Esh,99ML,06ML}. When the density of these quadrupoles is low, they act only to renormalize the elastic moduli, but they do not change the form of the theory. This is reminiscent the role of dipoles in dielectrics, where the dielectric constant is dressed, but the structure of electrostatic theory remains intact \cite{84LL}. On the other hand, when the density of quadrupoles is high, the gradients of their density cannot be
ignored, and these are acting as dipoles. Dipole-dipole and dipole-displacement interaction become crucial, and these change the structure of the theory and the resulting mechanical responses. 

{\bf Brief review of the theory:} To introduce the theory in its simplest form
we focus here on 2-dimensional systems with radial geometry. Specifically, we consider
amorphous configurations of binary disks, with and without friction. For the frictionless case we examined in numerical simulations the displacement field that results from an inflation of a disk closest to the center. For the frictional case our collaborators studied experimentally the displacement field that is induced by a pusher at the center of coordinates whose effect is similar to the inflation considered {\it in silico}. 
Details of these simulations and experiments can be found in Refs.~\cite{21LMMPRS,21MMPRSZ}. In short,  in our experiment and simulations we consider an annulus of radii $\rin$ and $\rout$, $\rin\ll \rout$, with an imposed displacement $\mathbf{d}(\rin) = d_0 \hat{r}$ and $\mathbf{d}(\rout) = 0$. The polar symmetry of the problem implies that $\mathbf{d}(r) =d_r(r) \hat{r}$, in which case normal elasticity theory implies the equation
\begin{equation}
	\Delta {\mathbf{d}}=	d_r'' +\frac{d_r'}{r}  -\frac{d_r}{r^2}=0 \ .
	\label{usual}
\end{equation}
The solution to this differential equation that satisfies the boundary conditions is
\begin{equation}
	{d}_r(r)=d_0 \frac{r^2 - \rout^2}{\rin^2 - \rout^2}\frac{\rin}{r} \ .
	\label{renelas}
\end{equation}
One should note that the solution (\ref{renelas}) is a positive monotonically decreasing function of $r$, and it decays at large values of $r$ like $1/r$, as is expected in standard elasticity theory. It was shown \cite{21LMMPRS} that this form of the displacement field is expected to
remain valid also when there exists a low (or uniform) density of plastic events, although the coefficients may get renormalized. We then refer to this situation as ``quasi-elastic".

The situation changes qualitatively when the density of plastic events gains sizeable gradients, and these act as dipole sources \cite{21LMMPRS}. The screening caused by dipoles changes Eq.~(\ref{renelas}) to read
\begin{equation}
	d_r'' +\frac{d_r'}{r}  -\frac{d_r}{r^2}= -\kappa^2 d_r  \,
	\label{new}
\end{equation}
with $\kappa$ being an emergent constant that is not known a-priori. Below we refer to $\kappa$ as the ``screening parameter". Eq.~(\ref{new}) is equivalent to the Bessel equation whose solution, 
satisfying $d_r(r_{\rm in})=d_0$, $d_r(r_{\rm out})=0$, reads
\begin{equation}
	d_r(r)  = d_0 \frac{ Y_1(r \, \kappa ) J_1(r_\text{out} \kappa )-J_1(r \, \kappa ) Y_1(r_\text{out} \kappa )}{Y_1( r_\text{in} \kappa ) J_1(r_\text{out} \kappa )-J_1(r_\text{in} \kappa ) Y_1(r_\text{out} \kappa )} \ .
	\label{amazing}
\end{equation}
Here $J_1$ and $Y_1$ are the Bessel functions of the first and second kind respectively. For very small values of $\kappa$ the solution reduces back to \eqref{renelas}. Importantly, depending on the precise value of $\kappa$, Eq.~(\ref{amazing}) may non monotonic, negative, and even oscillatory. We refer to situations that agree
with the solution Eq.~(\ref{amazing}) and are either non-monotonic or oscillating as ``anomalous elasticity". 
Until now the numerical values of $\kappa$ had to be determined by fitting
Eq.~(\ref{amazing}) to experiments and simulations \cite{21LMMPRS,21MMPRSZ}. One consequence of the present Letter will be a direct determination of the $\kappa$ as shown below. 

\begin{figure}
	\hskip -1 cm
		\includegraphics[width=0.9\linewidth]{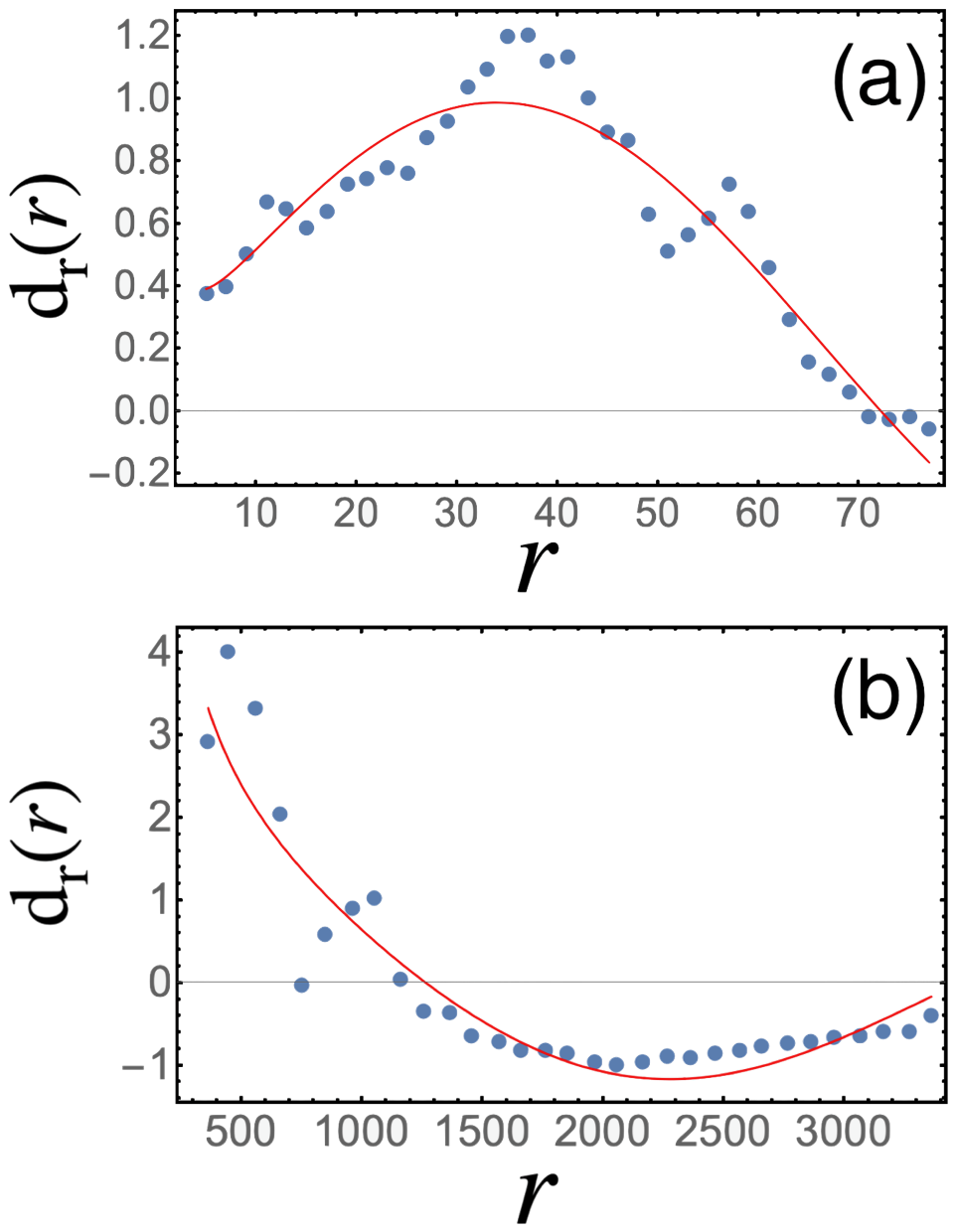}	
	\caption{Angle averaged radial components of the displacement field as measured
	in simulation and experiment, cf. Refs.~\cite{21LMMPRS,21MMPRSZ}. Panel (a): simulations with frictionless Hertzian disks, exhibiting an anomalous response. The continuous line is the solution Eq.~(\ref{amazing}) with $r_{\rm in}=4.8$, $r_{\rm out}=72.2$ and $\kappa=0.0525$. Panel (b): experimental measurements with frictional disks, The continuous line is  the solution Eq.~(\ref{amazing}) with  $r_{\rm in}=133.3$, $r_{\rm out}=3478$ and $\kappa=0.00147$. }
	\label{examples}
\end{figure}

{\bf Two examples for further analysis:} Our next aim is to establish the existence (or nonexistence) of dipole fields, 
directly from the measured displacement field.
We choose two examples, the first being a numerical simulation of frictionless binary amorphous assembly of disks \cite{21LMMPRS}, and the second an experiment with frictional binary disks \cite{21MMPRSZ}, all in circular geometry, and both showing anomalous responses. 
The displacement fields in these cases result from an inflation of a disk or a pusher
at the center of coordinates. The angle averaged radial component of the displacement field as measured in a simulation and an experiment respectively are shown in Fig.~\ref{examples}.
Both examples deviate strongly from the elastic solution (\ref{renelas}). Rather, the
continuous lines in both panels represent the anomalous solutions (\ref{amazing}) with $\kappa$ being a fitting parameter. Panels (a) and (b) are obtained with $\kappa=0.0525$ and $\kappa=0.00147$ respectively. 
We note here that the systems are of different sizes and $\kappa$ is dimensional. For the sake of comparison one should consider a dimensionless number $\tilde \kappa\equiv \kappa r_{\rm out}$. Then the two examples have $\tilde \kappa = 3.79$ and 5.11 respectively. 
One of our goals below
is to develop a direct measure for the screening parameter $\kappa$ and to test its success in predicting the curves in Fig.~\ref{examples}, without fitting.

{\bf Theory-independent measurement of dipoles:} We firstly consider the existence
of dipole sources in our measured displacement field, independently of our proposed
theory. To this aim we adopt the stress function formalism in which screening charges are introduced as sources for the bi-harmonic equation for the Airy function \cite{53Mus}. Distributed dipoles, either induced or fixed, are incorporated into the equation as a source term \cite{moshe2015elastic,irvine2010pleats}
\begin{equation}
	\frac{1}{Y}\Delta \Delta \chi =\nabla \cdot \Pvec
	\label{bihar}
\end{equation}
where $Y$ is the Young modulus and  $\Pvec(x)$ is the position dependent dipole density.

Remembering that the pressure $p$ is given by  $p \equiv \Delta \chi$, we
now consider the integral of the LHS of Eq.~\ref{bihar} over a domain $\Omega$: 
\begin{equation}
	\begin{split}
		\frac{1}{Y} \int _{\Omega }\Delta \Delta \chi  \text{dS} = \frac{1}{Y}\oint _{\partial \Omega } (\nabla \Delta \chi)\cdot \mathbf{n} \text{dl} = \frac{1}{Y}\oint _{\partial \Omega } (\nabla p) \cdot \mathbf{n}\, \text{dl}
	\end{split}
\end{equation}
where $\B n$ is the outgoing unit vector on the boundary of the domain. On the other hand, the integral of the RHS of Eq.~(\ref{bihar}) reads
\begin{equation}
		\int _{\Omega } \nabla\cdot \Pvec   \text{dS}  = 
	 \oint _{\partial \Omega }\Pvec\cdot \mathbf{n} \, \text{dl} \ .
\end{equation}
We therefore conclude that the  signature for the existence of non-uniform dipole distribution is from the integral over the pressure flux
\begin{equation}
	\begin{split}
		\frac{1}{Y}\oint _{\partial \Omega } (\nabla p) \cdot \mathbf{n}\, \text{dl} =  \oint _{\partial \Omega }\Pvec\cdot \mathbf{n} \, \text{dl}\;.
	\end{split}
\end{equation}
Note that the pressure is proportional to the trace of the strain $p = Y\, \mathrm{Tr}(u)$, which in turn describes the volume deformation of the material and is equal to the displacement divergence $\mathrm{Tr}(u) = \nabla \cdot \mathrm{d}$. Denoting the resulting integral as $I_1$ we end up with  
\begin{equation}
 \oint _{\partial \Omega }\Pvec\cdot \mathbf{n}	\, \text{dl}= \oint _{\partial \Omega } (\nabla ( \nabla \cdot \mathrm{d})) \cdot \mathbf{n}\, \text{dl} \equiv I_1  \;.
 \label{P}
\end{equation}
In the case of our circular systems with radial symmetry both sides of this equation
can be evaluated analytically.
The RHS reads
\begin{equation}
	\oint _{\partial \Omega } (\nabla ( \nabla \cdot \mathrm{d})) \cdot \mathbf{n}\, \text{dl} = 2\pi r \Big(d_r^{''}(r) +\frac{d_r'(r)}{r} -\frac{d_r(r)}{r^2}\Big)  \ ,
	\label{radial}
	\end{equation}
and the LHS reads
\begin{equation}
	\oint _{\partial \Omega }\Pvec\cdot \mathbf{n} = 2 \pi r P(r) \ ,
	\label{radialP}
\end{equation}
hence
\begin{equation}
	P(r)=  d_r^{''}(r) +\frac{d_r'(r)}{r} -\frac{d_r(r)}{r^2}  \ .
	\label{dipole1}
\end{equation}

We should notice that Eq.~(\ref{usual}) guarantees that Eq.~(\ref{radial}) is zero identically when classical elasticity is obeyed. Since this is correct for every closed circle, it means that $\B P(r)=0$. This is consistent with the absence
of dipole sources. Note also that in 2-dimensions the elasticity theory solution for the radial component of the displacement field, being proportional to $1/r$ at large values of $r$ results in a trivial zero on the RHS of Eq.~(\ref{radial}).

\begin{figure}
	\includegraphics[width=0.9\linewidth]{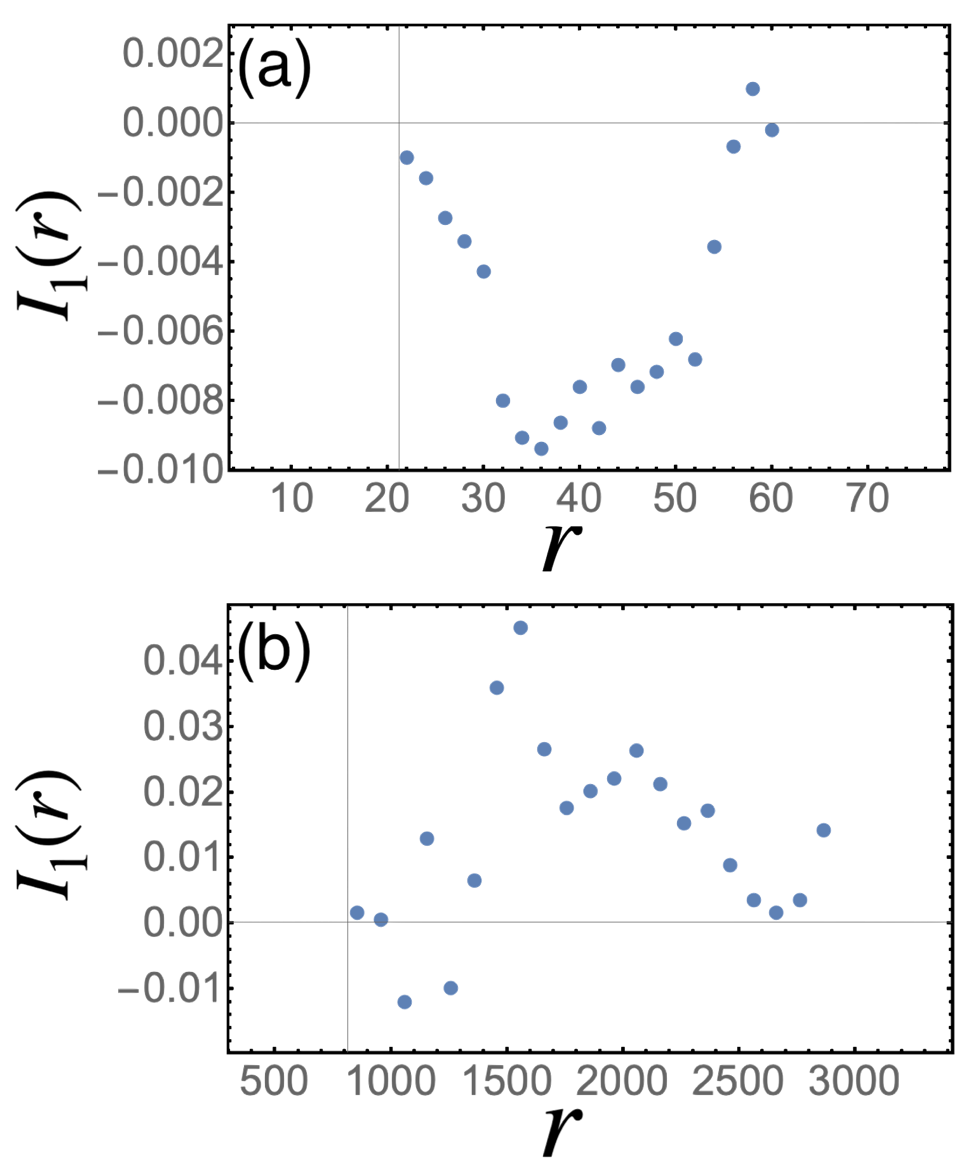}	
	\caption{The integral $I_1$ of Eq.~(\ref{radial}) computed for the two anomalous examples shown in panels (a) and (b) of Fig.~\ref{examples}. These plots use the data without any resort to theoretical fits. The existence of these integrals are a direct demonstration for the presence of dipole fields. }
	\label{dipoleyesh}
	\end{figure}

At this point we return to the two examples shown in Fig.~\ref{examples}, and compute the dipole density in Eq.~(\ref{dipole1}), directly from
the measured radial component. The results for the frictionless simulation and the frictional experiment  are shown in Fig.~\ref{dipoleyesh}.
\begin{figure}
	\hskip -0.3 cm
	\includegraphics[width=0.9 \linewidth]{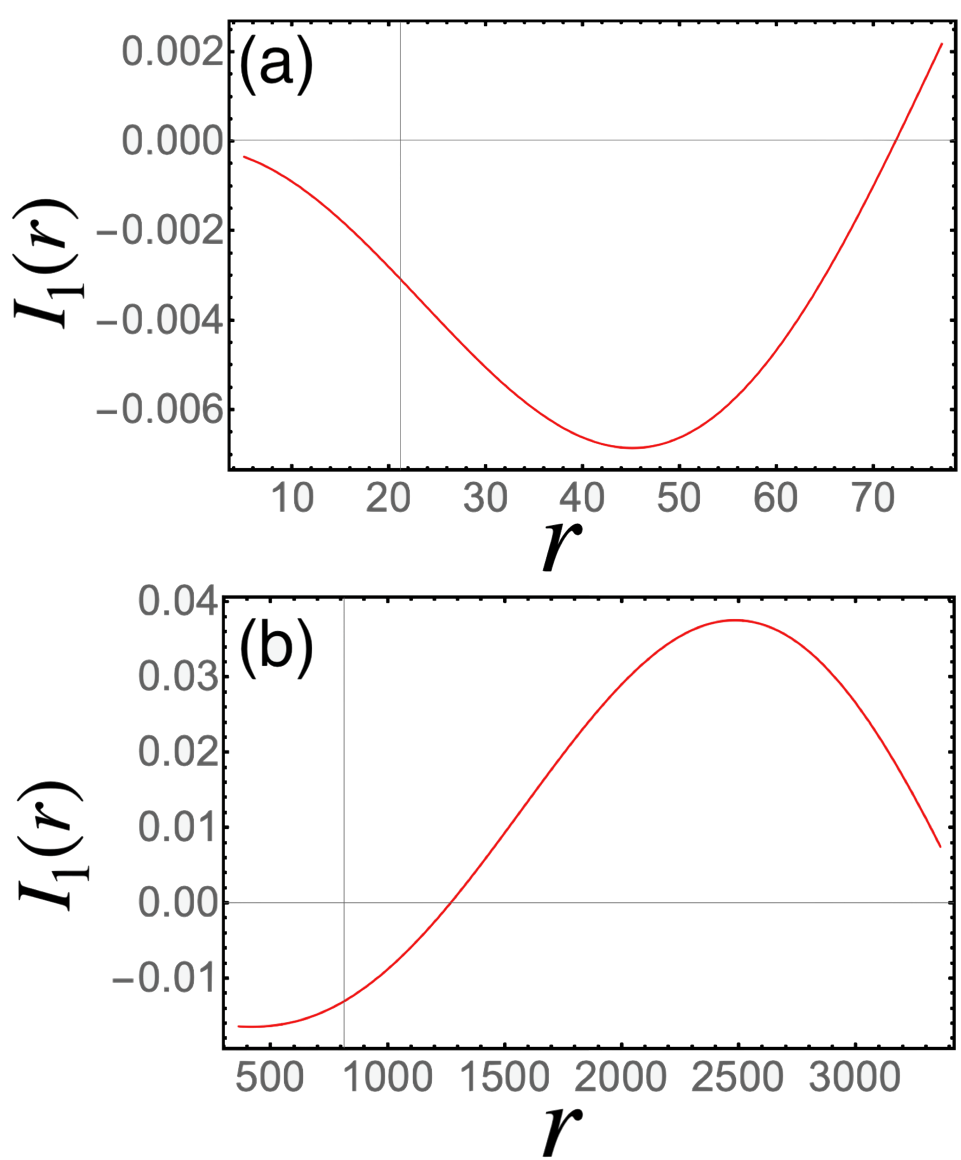}	
	\caption{The same integral $I_1$ of Eq.~(\ref{radial}) computed for the two examples
		shown in Fig.~\ref{examples}. Here we use the fit Eq.(\ref{amazing}) for the sake of smoothness. These integrals are identical to those obtained using Eq.~(\ref{int2}) up to the scale $\kappa^2$ as is demanded from the solution Eq.~(\ref{amazing}).}
	\label{I2}
\end{figure}		
Although the data is somewhat noisy, we can see that the measured dipole density is non-zero for the two cases, indicating that indeed dipole sources exist in the system, as they emerge together with the onset of the displacement fields. We stress that this result is independent of the theory proposed
in Ref.~\cite{21LMMPRS}. To obtain a smoother measurement we use the fit of the data to Eq.~(\ref{amazing}). The result of this smoother evaluation is shown for the two examples in Fig.~\ref{I2}. The two methods of evaluation are consistent with each other and show that a dipole
field is emerging together with the displacement field which is caused by the inflation near
the center of our amorphous solids in their radial geometry.

{\bf Theory dependent evaluation:} Next we test the theory that was proposed in Ref.~\cite{21LMMPRS}
where an anomalous response of the system is described by Eq.~(\ref{new}) 
\begin{equation}
	\nabla \mathrm{Tr}(u) = (\nabla ( \nabla \cdot \mathrm{d})) =-\kappa^2 \mathbf{d}
\end{equation}
hence the integrand in LHS of Eq.~(\ref{radial}) can be simplified to
\begin{equation}
		\oint _{\partial \Omega }\Pvec\cdot \mathbf{n} \, \text{dl} = -\kappa^2 \oint _{\partial \Omega }\mathbf{d}\cdot \mathbf{n} \, \text{dl} \ .
		\label{test}
\end{equation}
If the theory is correct, the evaluation using the RHS of Eq.~(\ref{test}) should
agree with the evaluation of Eq.~(\ref{radial}). Denote the integral itself as $I_2$,
\begin{equation}
	I_2\equiv -\oint _{\partial \Omega }\mathbf{d}\cdot \mathbf{n} \, \text{dl}\ .
	\label{int2}
\end{equation}
Since we have used in Fig.~\ref{I2} the solution of Eq.~(\ref{amazing}) to compute $I_1$, it is obvious that $I_2$ would be identical to $I_1$ up to the scale factor $\kappa^2$. 
Computing  $\sqrt{I_1/I_2}$ one indeed finds the evident value of $\kappa$ which was used to fit the data in Fig.~\ref{examples}.

\begin{figure}
	\vskip 0.4 cm
	\includegraphics[width=0.9\linewidth]{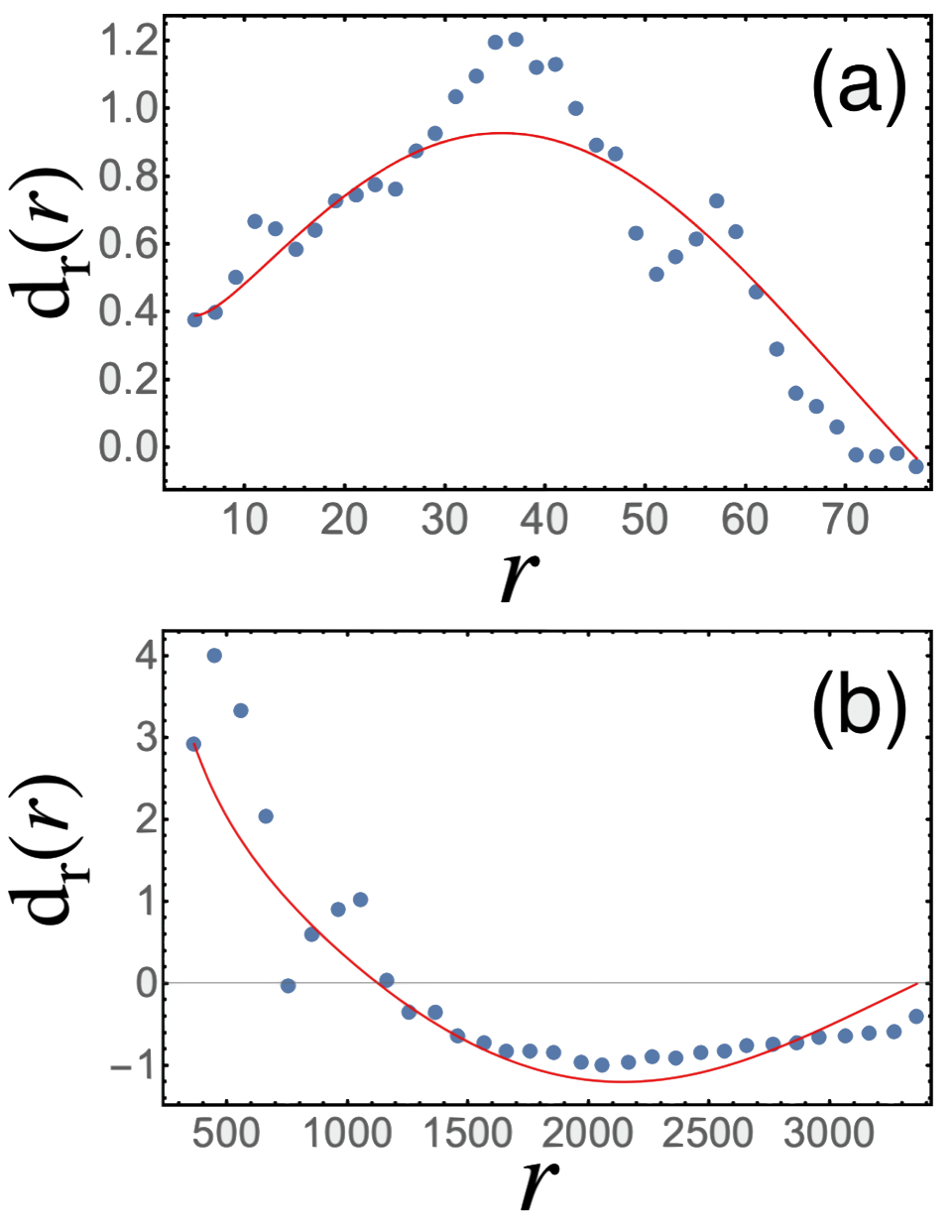}	
	\caption{Comparison of the solutions obtained using the screening parameter $\kappa$ from
		the measurements of the dipole fields. Panel (a): comparison to the experimental data (blue points) as shown in Fig.~\ref{examples} panel (a). Panel (b) comparison to the simulation data (blue points) shown in panel (b) of  Fig.~\ref{examples}.}
	\label{inverse}
\end{figure}		

Thus so far we did not put the theory of Ref.~\cite{21LMMPRS} under a stringent test.
To achieve such a test we return to the raw data of the angle averaged displacements
shown in Fig.~\ref{examples}. To avoid effects of noise we smooth the displacement data by calculating moving averages. The displacement at each position $r$ is estimated from the average of 8 nearest neighbors.  In addition, we note
that the integrals $I_{1,2}$ measure the dipole {\em density} $P(r)$. Upon integrating over the radial direction we find the total dipole charge induced in the system
\begin{equation}
	\begin{split}
		\mathcal{P}_\mathrm{tot} & = 2\pi \int r\, P(r) \mathrm{d}r  \\
		&= \kappa^2 \int  I_2(r) \mathrm{d}r =  \int  I_1(r) \mathrm{d}r
	\end{split}
\label{Ptot}
\end{equation}
Therefore we can estimate $\kappa$ by
\begin{equation}
	\begin{split}
		\kappa =\sqrt{ \frac{ \int  I_1(r) \mathrm{d}r}{ \int  I_2(r) \mathrm{d}r} }\ .
	\end{split}
\label{kappa}
\end{equation}
This ratio can be calculated directly from the raw data. We find the values  $\kappa=0.04980$ and $\kappa= 0.00146$ for the two examples in Fig.~\ref{examples}. If the theory is correct, when these measured calues of $\kappa$ are plugged in to the predicted form in Eq.~(\ref{amazing}) the resulting curves should agree with the measured radial components of the displacement without performing any fitting procedure. 
Indeed, in Fig.~\ref{inverse} we plot the predicted form (\ref{amazing}) and the measured displacement using  the calculated $\kappa$ and find very good agreement, thus providing a support to the dipole screening picture in general and to our specific model in particular.

{\bf Summary and conclusions:} The aim of this Letter was twofold. Firstly we wanted to present a direct computation of the dipole field that was presumed to exist in Refs.~\cite{21LMMPRS,21MMPRSZ}. Contrary to the case of electrostatics \cite{84LL}, where dipoles can be directly attributed to polarized molecules or polymers, in the present case dipoles were identified as gradients of quadrupole fields, rendering them less obvious. We have shown in this Letter how to compute the dipole fields from
loop integrals on the displacement field. The existence of a dipole field goes hand in hand with a finite value of the screening parameter $\kappa$ and with having anomalous elasticity. {\it Mutatis mutandis}, when $\kappa=0$ we expect quasi-elastic responses with a possible renormalization of the elastic moduli. 

The second aim of this Letter was to present a method to estimate the numerical value of the screening parameter $\kappa$ from the measurement of the dipole field.
The total dipole charge, as defined in Eq.~(\ref{Ptot}), provides us with a method
to compute $\kappa$ from Eq.~(\ref{kappa}), without fitting the angle-averaged displacement field. The agreement of the prediction of Eq.~(\ref{amazing}) with the data, once we used the screening parameter from this procedure, is a strong support for the theory as presented in Refs.~\cite{21LMMPRS,21MMPRSZ}.

It should be stated that until now the theory was tested by and compared to simulations and experiments involving disks in radial geometry in 2-dimensions. In the near future we will provide evidence that the theory is also applicable to systems of spheres in 3-dimensions, and to standard models of glass formers in 2 and 3-dimensions. 

{\bf acknowledgments}: We thank Chandana Mondal for helpful discussions and sharing data. This work had been supported in part by Minerva Foundation and the Minerva Center for ``Aging, from physical materials to human tissues" at the Weizmann Institute.  MM acknowledges support from the Israel Science Foundation (grant No. 1441/19).

\bibliography{friction.anomalous}

\end{document}